\documentclass[10pt,prd,twocolumn,showpacs,preprintnumbers,amsmath,amssymb]{revtex4-1}
\usepackage[utf8]{inputenc}
\usepackage[T1]{fontenc}
\usepackage{graphicx}
\usepackage{amsmath}
\usepackage{amsfonts}
\usepackage{amssymb}

\begin{document}
\newcommand{\wthj}[6]{
\setlength{\arraycolsep}{2pt}
\left(\begin{array}{ccc}
#1 & #3 & #5\\#2 & #4 & #6                                                                                        
\end{array}\right)
\setlength{\arraycolsep}{6pt}
}
\newcommand{\thjr}[2]{
\sqrt{#1\!\cdot\!(2l\!+\!1)\!\cdot\! #2}
}
\newcommand{\ylm}{{Y^l_m}^\ast(\hat{q})}
\newcommand{\lsum}[1]{\sum_{l=0}^#1\sum_{m=-l}^l}
\newcommand{\pipi}{$\pi\pi$ }
\title{Tensor meson photoproduction as a final state interaction effect}
\author{Łukasz Bibrzycki}
\email[Corresponding author.\\]{lukasz.bibrzycki@ifj.edu.pl}
\affiliation{State School of Higher Education in Oświęcim, Kolbego 8, 32–600 Oświęcim, Poland}
\author{Robert Kamiński}
\email[Corresponding author.\\]{robert.kaminski@ifj.edu.pl}
\affiliation{Institute
of Nuclear Physics, Polish Academy of Sciences, Division of Theoretical Physics, 31-342 Kraków, Poland}
\begin{abstract}
 The model is presented to describe the $f_2(1270)$ meson photoproduction as a result of pion-pion interactions in the final state. 
Treating tensor mesons as objects dynamically created due to final state interactions is a convenient and 
straightforward way to employ data from $\pi\pi$ scattering like phase shifts and inelasticities for description of 
(photo)production reactions while retaining proper analytical structure of amplitudes, two particle unitarity and 
crossing symmetry. The model presented here can provide experimentally testable quantities like differential cross 
sections 
and  $\pi\pi$ mass distributions as well as the strengths of partial waves corresponding to various $f_2(1270)$ helicities 
which are essential for partial wave analyses. It can also be used to compute moments of angular distribution and spin density matrix elements where partial wave interference effects are important.
\end{abstract}
\pacs{13.60.Le, 13.75.-n, 13.60.-r, 14.40.-n}
\maketitle
\section{Introduction}
Description of the spectrum of resonances observed in the $\pi\pi$ (and $K\overline{K}$) system and excited in photon nucleon 
collisions is one of the most challenging problems of hadron spectroscopy. In the diffractive region of high energies and 
low momentum transfers, this reaction is dominated by vector meson production generated by Pomeron exchange, and its theory is 
quite firm \cite{Pump,DonLan,Pichow}. In the lower 
energies, the $P$-wave $\pi^+\pi^-$ photoproduction was described in terms of the $t$-channel exchange of Reggeons 
\cite{FriSoy,OhLee}. Attempts have also been made to include the intermediate nucleon resonances through various $s$-channel 
and $u$-channel mechanisms \cite{TejeOset1,TejeOset2,NachOset}. For the 
photoproduction of the $S$-wave and $D$-wave resonances, the situation is not clear both experimentally and theoretically. Because of
small photoproduction cross sections, they are very difficult to observe in mass distributions. So the method of choice is to
analyze the interference patterns of the weak $S$- and $D$-wave amplitudes with dominant $P$-wave amplitude. The partial wave 
interference can be conveniently analyzed with moments of pion angular distribution or spin density matrix elements.
 Such an approach was 
employed in a recent analysis of the reaction $\gamma p \rightarrow\pi^+\pi^-p$ performed by the CLAS group at Jefferson 
Laboratory, where the first observation of $f_0(980)$ photoproduction was reported 
\cite{CLAS}. The same experiment saw the $f_2(1270)$ signal, which previously was also observed by Hermes 
experiment at Deutsches Elektronen-Synchrotron using similar 
methods \cite{Hermes}. The apparent sensitivity of moments analysis in the search for a signal of rare resonances has a 
reverse, however, namely, that it requires proper accounting for all relevant production mechanisms. Nevertheless 
this method has been successfully employed to extract the $f_0(980)$ and $a_0(980)$ from the photoproduced $K\overline{K}$ 
spectrum \cite{BibLesSzcz} and $f_0(980)$ from the $\pi^+\pi^-$ spectrum \cite{BibLes}. The amplitude of $f_2(1270)$  
photoproduction is the necessary ingredient in order to properly describe the partial wave interference pattern for {\bf $\pi\pi$}
effective masses above 1~GeV. 

Previously, the electromagnetic processes involving tensor mesons were discribed in terms of the 
combined tensor meson dominance and vector meson dominance models \cite{OhLee,Renner,Suzuki}, Regge inspired exchange models
\cite{Trian,Ahma}, or effective field theories \cite{Toublan,Giacosa}. None of these approaches can, however, be 
treated as properly tested in tensor meson photoproduction on a nucleon. 
Production of $f_2(1270)$ has been extensively analyzed in other reactions like
$\gamma \gamma\to\pi^+\pi^-$ and $\gamma \gamma\to\pi^0\pi^0$ \cite{Achasov1,Achasov2}. The authors of these studies found that this
resonance is dominantly produced in quark-antiquark channel and that pion-pion
final state interactions are negligible. We note, however, that qualitative characteristics of the $\gamma\gamma\to\pi\pi$
reaction are quite different from those of $\gamma p \to \pi^+\pi^- p$ photoproduction.
For example, the $f_0(980)$ signal which is relatively small yet clear in
$\gamma\gamma\to\pi\pi$ reaction analyzed by Belle \cite{Belle} is completely absent in $\pi^+\pi^-$
mass distribution of $\gamma p \to \pi^+\pi^- p$ reaction measured by CLAS, even
though the data errors and mass resolution of 10 MeV are in principle sufficient 
to observe it (see eg. Fig. 4 of \cite{CLAS}). It was only due to $f_0(980)$
interference with the dominant $P-$wave that the $f_0(980)$ has been observed. On the other hand, in the $\pi^+\pi^-$ and $\pi^0\pi^0$ mass 
distributions measured by Belle Collaboration \cite{Belle}, the $f_2(1270)$ signal hugely outnumbers the $f_0(980)$ one. This is in contrast 
with the mass distributions of the $S-$ and $D-$waves measured by CLAS and integrated in the neighborhood of the $f_0(980)$ and $f_2(1270)$ resonances  for the
$\gamma p \to \pi^+\pi^- p$ reaction. In this measurement, the $f_0(980)$ cross section is smaller than the $f_2(1270)$ one only by a factor of about 4 (see figs. 22 and 24 of \cite{CLAS}). We also stress that our approach does not contradict the $q\bar{q}$ nature of the $f_2(1270)$ resonance. This is because the 
$\pi\pi$ amplitudes which we use as the input were derived in the model independent way. Moreover, our assumptions concerning the meson exchanges included 
in Born amplitudes can be directly checked by comparison of model predictions with precision data on partial wave interference in the $\pi\pi$ effective mass range corresponding to the $f_2(1270)$ resonance.

In what follows, we will refer to $\pi^+\pi^-$ 
photoproduction as description of $D$-wave data from CLAS is our main objective. One has to mention, however, that the formalism we 
present is \textit{mutatis mutandis} applicable to $\pi^0\pi^0$ photoproduction.

\section{Model for the {\boldmath$ \pi^+\pi^-$} photoproduction}
\subsection{Born amplitudes}
\begin{figure}[h]
 \centering
 \includegraphics[scale=0.66,clip]{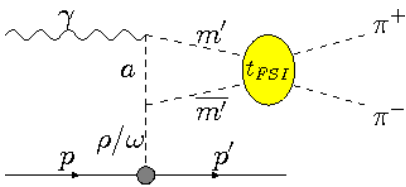}
 \caption{The diagram of two pion photoproduction with final state interactions where $a$ denotes $\pi$, $\rho$ or $\omega$.}
 \label{fig:1}
\end{figure}

\begin{figure}[h]
 \centering
 \includegraphics[bb=20 15 150 332, scale=.9, clip]{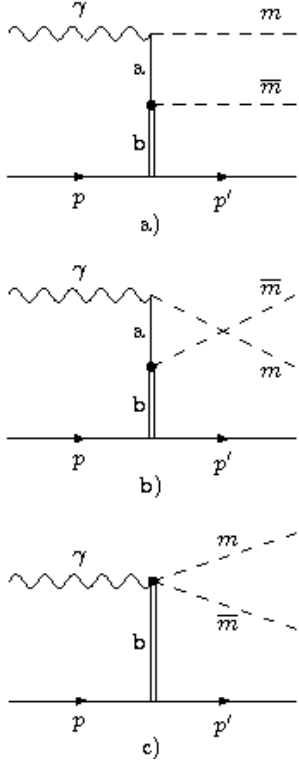}
 \caption{The structure of diagrams corresponding to Born photoproduction amplitudes}
 \label{fig:born}
\end{figure}

In our approach the tensor meson photoproduction is treated as a two-stage phenomenon. First, a pair of pions is 
photoproduced. According to Regge phenomenology, this process at high energies should be dominated by $t$-channel $\rho$ and 
$\omega$ exchanges. Then pions undergo the final state interactions which may result in the resonance creation. This two-stage process is
 schematically drawn in Fig.\ref{fig:1}. The principal merit of the model we propose is that it preserves important 
features of the
$\pi\pi$ scattering amplitudes described in Sec. \ref{ampfsi} like two-particle unitarity, proper analytical structure, and
crossing symmetry and embeds them seamlessly in the framework of the photoproduction amplitude.
We follow the general formalism of Refs. \cite{JiKaLe,LeSzcze} but specialize the results to the 
case of two pions photoproduced in the $D$-wave (for completeness we will recall some important formulas of these references).
The amplitude of the final state interactions is described in \cite{GKPYIV,KDF}.
In principle, our approach does not engage any new parameters, as coupling constants and form factor range parameters are common for all partial waves and the same as in \cite{JiKaLe}. The vector meson to nucleon couplings are taken from Bonn model \cite{Bonn} and so is
the monopole form factor used in $VNN$ vertex. In practice,
however, the cross sections computed with these parameters substantially overestimate the experimentally measured data. So we 
leave ourselves with the freedom to use the overall rescaling factor to adjust the cross section predicted by the model to experimental data. 
Thus, we treat the relative strengths
of partial waves corresponding to different angular momentum projections as the principal model prediction. These depend mainly on the 
meson exchanges taken into account in the model. We believe that any possible variations of couplings will not change the picture presented here substantially. Predictions
for partial wave amplitude strengths (and phases) are important components for analysis of moments of $\pi^+\pi^-$ angular
distribution. This analysis will be discussed in the paper to follow \cite{BiKa} with application of amplitudes discussed here.
Our calculations are performed in the helicity system which is the center of the mass system of the two photoproduced 
pions. In this system the z-axis is directed opposite to final proton momentum
$\boldsymbol{p}'$, the $y$-axis is perpendicular to the production plane, and the $x$-axis versor is defined as $\hat{x}=\hat{y}\times\hat{z}$.
We describe the initial state $\pi\pi$ photoproduction in terms of Born amplitudes derived from the phenomenological 
Lagrangian:
\begin{equation}
\begin{split}
 \mathcal{L}&=
\mathcal{L}_{\pi\pi\gamma}+\mathcal{L}_{\rho\pi\gamma}+\mathcal{L}_{\omega\pi\gamma}+\mathcal{L}_{\rho\pi\pi\gamma}\\&+
\mathcal{L}_{\rho\pi\pi}+\mathcal{L}_{\rho\pi\omega}+
\mathcal{L}_{\omega NN}+\mathcal{L}_{\rho NN},
\label{lagr}
\end{split}
\end{equation}
where individual terms of Eq.(\ref{lagr}) are defined in \cite{JiKaLe}. The diagram representation of amplitudes obtained 
from this Lagrangian is shown in Fig.\ref{fig:born} 
and they have a general form of
\begin{equation}
 V_{m\overline{m}}=\sum_{r=I,II}\overline{u}(p',s')J_{r,m\overline{m}}\cdot\varepsilon(q,\lambda_\gamma)u(p,s),
 \label{fullamp}
\end{equation}
where $J_{r,m\overline{m}}$ is the hadronic current, $u(p,s)$ and $\overline{u}(p',s')$ - wave functions of the initial and 
final proton, respectively, and $\varepsilon$ the polarization vector of the incident photon which reads
\begin{equation}
 \varepsilon(q,\lambda_\gamma)=(0,\boldsymbol{\varepsilon}^{\lambda_{\gamma}}),
\label{polvec}
\end{equation}
where
\begin{equation}
 \boldsymbol{\varepsilon}^{\lambda_{\gamma}}=-\frac{\lambda_\gamma}{\sqrt{2}}(\cos\theta_{\!q},i\lambda_\gamma,\sin\theta_{\!q})
\end{equation}
and $\lambda_\gamma$ is photon helicity.
The photon polarization vector is transverse to photon momentum:
\begin{equation}
 \boldsymbol{q}=|\boldsymbol{q}|(-\sin\theta_{\!q},0,\cos\theta_{\!q}),
\label{photmom}
\end{equation}
and
\begin{equation}
 \cos\theta_{\!q}=\frac{E^2-{E'}^2-|\boldsymbol{q}|^2}{2|\boldsymbol{q}||\boldsymbol{p}'|}.
\end{equation}
The energies $E$ and $E'$ of the initial and final proton respectively, as well as photon energy $|\boldsymbol{q}|$ can be 
expressed in terms of Lorentz invariant quantities:
\begin{equation}
 E=\frac{s-m^2+t}{2M_{\pi\pi}}, \qquad E'=\frac{s-m^2-M_{\pi\pi}^2}{2M_{\pi\pi}},
\end{equation}
\begin{equation}
 |\boldsymbol{q}|=\frac{M_{\pi\pi}}{2}-\frac{t}{2M_{\pi\pi}},
\end{equation}
where $s$ is the $\gamma p$ energy squared, $t$ is the square of the 4-momentum transfer from initial photon 
to the photoproduced $\pi\pi$ system, $m$ is the proton mass and $M_{\pi\pi}$ is the effective mass of two pions.

In Eq.(\ref{fullamp}) $r=I$ corresponds to the sum over diagrams where $a=\pi$ in Fig.\ref{fig:born} (including the 
contact diagram) and $r=II$ corresponds to the sum of diagrams with $a=\rho$ or $\omega$. The summary of these diagrams is 
shown in Table~\ref{tabdia}.
\begin{table}[h]
\begin{center}
\begin{tabular}{lcc}
\hline
$m\overline{m}$ & r=I & r=II \\ 
\hline
$\pi^+\pi^-$ &$(a,b)=(\pi^{\pm},\rho^0)$   & $(a,b)=(\rho^{\pm},\omega)$ \\ 
$\pi^0\pi^0$ &  & $(\rho^{0},\omega)$, $(\omega,\rho^0)$\\
\hline
\end{tabular}
\caption{Summary of meson exchanges in Born amplitudes.}
\label{tabdia}
\end{center}
\end{table}
The amplitude defined in Eq.(\ref{fullamp}) is then $D$-wave projected using the formula:
\begin{equation}
  V_{m\overline{m}}^{2M}=\frac{1}{\sqrt{4\pi}}\int d\Omega {Y^2_M}^\ast(\Omega) V_{m\overline{m}}.
 \label{pwe}
\end{equation}
In our frame of reference, the momenta of photoproduced pions can be 
expressed in terms of the solid angle $\Omega$, i.e., $\boldsymbol{k_1}=-\boldsymbol{k_2}=|k|\hat{\kappa}(\Omega)$. 
$\boldsymbol{k_1}$($\boldsymbol{k_2}$) is the positive (negative) pion momentum and $\hat{\kappa}=(\sin\theta 
\cos\varphi,\sin\theta\sin\varphi,\cos\theta)$. In what follows we will write just $k$ instead of $|k|$ for brevity.
The general form of the current used in Eq.(\ref{fullamp}) is
\begin{equation}
\begin{split}
 J^{\mu}_{r,m\overline{m}}&=(\alpha_{r,m\overline{m}}g^{\mu\nu}+k_1^{\mu}\beta_{1r,m\overline{m}}^\nu+
k_2^\mu\beta_{2r,m\overline{m}}^\nu)\\
&\quad\times\lbrace d_{r,m\overline{m}}\gamma_\nu+e_{r,m\overline{m}}(p+p')_\nu\rbrace
\end{split}
\label{current}
\end{equation}
where functions $\alpha_{r,m\overline{m}}$, $\beta_{1r,m\overline{m}}$, $\beta_{2r,m\overline{m}}$, $d_{r,m\overline{m}}$  and
$e_{r,m\overline{m}}$ are defined in \cite{JiKaLe}.
It is interesting to note that terms of Eq.(\ref{current}) contained in curly braces do not depend on pion momenta, and thus they 
can be factorized out of the partial wave expansion. Physically, it means that in this model the $D$-wave angular momentum 
projections $M$ are uncorrelated with nucleon spin projections. Finally, after all pion momentum independent terms are 
factorized out of Eq.(\ref{pwe}) we arrive at the $D$-wave projected tensor defined as
\begin{equation}
\begin{split}
 P_{r,m\overline{m}}^{2M,\mu\nu}=&\frac{1}{\sqrt{4\pi}}\int d\Omega
 {Y^2_M}^\ast(\Omega)\\&(\alpha_{r,m\overline{m}}g^{\mu\nu}+k_1^{\mu}\beta_{1r,m\overline{m}}^\nu+
k_2^\mu\beta_{2r,m\overline{m}}^\nu).
\end{split}
\end{equation}
Because of the photon polarization vector definition [Eq.(\ref{polvec})], the only matrix elements of the tensor
 $P_{r,m\overline{m}}^{2M}$ which enter the amplitude are $P_{r,m\overline{m}}^{2M,i0}$ and $P_{r,m\overline{m}}^{2M,ij}$,
where $i,j=x,y,z$. We stress that the form of the tensor $P_{r,m\overline{m}}^{2M}$ is general and it can be used to construct other amplitudes to 
describe transition of two vector particles into two pseudoscalar ones, e.g. $\gamma\gamma^\ast\to m\overline{m}$, where 
$m\overline{m}$ can be $\pi\pi$, $K\bar K$, $\pi\eta$. Therefore the full expressions for individual matrix elements of the tensor $P_{r,m\overline{m}}^{2M}$ for $r=I$ and $r=II$ are given in the Appendix.
\subsection{Final state scattering amplitudes}\label{ampfsi}
The $\pi\pi$ final state scattering amplitudes $t_{\ell}^{I}(s_{\pi\pi}) =
t_{2}^{0}(s_{\pi\pi})$ and $t_{2}^{2}(s_{\pi\pi})$ for the $D$ wave with
isospin 0 and 2 respectively, have been
described using parameterization constructed and used in the recent dispersive
data analysis \cite{GKPYIV}.
Their advantage over other parameterizations is unitarity, analyticity and model
independent formalism.
The $D$-wave amplitudes have been fitted to experimental data up to 1.42 GeV and
indirectly to a system of dispersion relations below 1.1 GeV.
Two of these relations were the Roy like ones, i.e., relations with imposed crossing
symmetry condition.
One of them, presented and called for short GKPY in\cite{GKPYIV}, has been derived with one subtraction and proved to
be very demanding which allowed for very
precise determination of directly fitted amplitudes ($S$ and $P$) and,
indirectly, other ones ($D$, $F$ and $G$).
The general form of dispersion relations with one subtraction for the $D$-wave
amplitudes reads:
\begin{equation}
\begin{split}
  \mbox{Re } t_{2}^{I}(s_{\pi\pi})&= {  d_{2}^{I}(s_{\pi\pi})}\\
      &\displaystyle \hspace{-1.3cm}+\sum\limits_{I'=0}^{2}
      \displaystyle \sum\limits_{\ell'=0}^{3}
     \hspace{0.2cm}-\hspace{-0.75cm}\int\limits
\limits_{4m_{\pi}^2}^{{s'}_{\pi\pi;max}}\hspace{-0.4cm} d{s'}_{\pi\pi}
   {  K_{2 \ell^\prime}^{I I^\prime}(s_{\pi\pi},s_{\pi\pi}')} {\mbox{Im
}t_{\ell'}^{I^\prime}
   ({s'}_{\pi\pi})}
\end{split}
\label{DF_OSDR}
\end{equation}
where $I = 0,2$, $s_{\pi\pi} = M_{\pi\pi}^2$, ${s'}_{\pi\pi;max}$=1.42 GeV$^2$, and
the factors $K_{2 \ell^\prime}^{I I^\prime}(s_{\pi\pi},{s'}_{\pi\pi})$ are
kernels derived with an imposed crossing symmetry condition.
Terms $d_{2}^{I}(s_{\pi\pi})$ comprise contributions from all partial waves
above $s_{\pi\pi}' =
{s'}_{\pi\pi;max}$ where the input amplitudes are described by using
the Regge formalism.
Below ${s'}_{\pi\pi} = {s'}_{\pi\pi;max}$, all partial wave amplitudes
$t^{I'}_{\ell'}({s'}_{\pi\pi})$ are
parameterized by using simple polynomials for phase shifts
$\delta({s'}_{\pi\pi})$ and inelasticities $\eta({s'}_{\pi\pi})$ which
guarantees their unitarity; see \cite{GKPYIV} for details.
These amplitudes can be expressed by
experimental $\delta_{\ell'}^{I'}(s_{\pi\pi})$ and
$\eta_{\ell'}^{I'}(s_{\pi\pi})$:
\begin{equation}
t^{I'}_{\ell'}({s'}_{\pi\pi})=\frac{{s'}_{\pi\pi}(\eta_{\ell'}^{I'}(s_{\pi\pi})e^{i\delta_{\ell'}^{I'}(s_{\pi\pi})}-1)}
{2i\sqrt{{s'}_{\pi\pi}-4m_{\pi}^2}}.
\label{Eq:ampl}
\end{equation}  
For the isoscalar $D$ wave, these are of course dominated by the $f_2(1270)$ resonance.

As has been presented in \cite{KDF}, although the $D$-wave amplitudes were not
fitted directly to the GKPY dispersion relations, they very well fulfill  crossing
symmetry condition below about 0.8 GeV and quite well above this energy.

Another argument in favor of our choice of parameterization was that
(see \cite{GKPYIV,KDF}), although all amplitudes ($S$-$G$ partial
waves) have been fitted separately to their "own" data, they all had
to be related with each other in very wide energy range via simultaneous
fit to the $S$ and $P$ waves.
These mutual relations are due to theoretical crossing symmetry condition
imposed on the amplitudes in the Roy and GKPY equations.
It guarantees mutual
consistency of all partial wave amplitudes and allows to believe
that isoscalar $D$ wave amplitude will not need any sizable further
modifications in future.
\subsection{Complete photoproduction amplitudes}
The complete (i.e. including the final state interactions) amplitude of the $D$-wave $\pi^+\pi^-$ photoproduction contains the information on energy and momentum transfer dependence of $\pi\pi$ photoproduction as well as the pion momentum (or effective mass) dependence of the
$\pi\pi$ scattering amplitude with proper analytical structure encoded. It reads
\begin{widetext}
\begin{equation}
\begin{split}
\langle \lambda' M |A_{\pi^+\pi^-}|\lambda_\gamma \lambda \rangle&=
    \langle \lambda' M|\hat{V}_{\pi^+\pi^- }|\lambda_\gamma \lambda\rangle\\&+
    4\pi\!\!\sum_{m'\overline{m}'}\int_0^\infty\frac{{k'}^2 dk'}{(2\pi)^3}F(k,k')\langle \pi^+\pi^- 
    |\hat{t}_{FSI}|m'\overline{m}'\rangle  G_{m'\overline{m}'}({M'}_{\pi\pi})
    \langle \lambda' M|\hat{V}_{m'\overline{m}'}|\lambda_\gamma \lambda\rangle
\end{split}    
\label{photoamp}
\end{equation}
\end{widetext}
where $\hat{V}$ is the Born amplitude of the $\pi^+\pi^-$ or $\pi^0\pi^0$ photoproduction, $\hat{t}_{FSI}$ is the $\pi\pi$ scattering amplitude, $\lambda, \lambda', 
\lambda_\gamma$ and $M$ are, respectively, the helicities of the initial and final proton, photon helicity, and projection of
the $\pi\pi$ system angular momentum on the spin quantization axis $z$ [which can be identified with the $f_2(1270)$ helicity]. $\hat{G}$ is 
the propagator of the intermediate pion pair and reads
\begin{equation}
 G_{m'\overline{m}'}({M'}_{\pi\pi})=\frac{1}{M_{\pi\pi}-{M'}_{\pi\pi}(k')+i\varepsilon}.
\label{prop}
\end{equation}
$F(k,k')$ is the form-factor needed to regularize divergent mesonic loop of diagram shown in Fig.\ref{fig:1}. Results 
obtained in the $S$-wave calculations \cite{JiKaLe} suggest that the particular value of this form-factor cut-off parameter  
may strongly affect calculated cross sections, and thus it should be carefully fitted to the data. In this explanatory study we 
limit ourselves to the on-shell part of the amplitude and leave the problem of fitting the form-factor parameter for further
investigation. After integration and rewriting the $\pi\pi$ amplitude in terms of isospin amplitudes, we arrive at the 
following expression:
\begin{equation}
\begin{split}
 &\langle \lambda' M |\hat{A}_{\pi^+\pi^-}|\lambda_\gamma \lambda \rangle=\\& \big[1+ir_{\pi}\big(\frac{2}{3}t^{I=0}_{\pi\pi}+
\frac{1}{3}t^{I=2}_{\pi\pi}\big)\big]\langle \lambda' M|\hat{V}_{\pi^+\pi^-}|\lambda_\gamma \lambda\rangle\\&+
\frac{1}{3}\big[ir_\pi(-t^{I=0}_{\pi\pi}+t^{i=2}_{\pi\pi})\big]
\langle \lambda' M|\hat{V}_{\pi^0\pi^0}|\lambda_\gamma \lambda\rangle,
\end{split}
\label{amplD}
\end{equation}
where $r_{\pi}=-kM_{\pi\pi}/8\pi$. First term in Eq.(\ref{amplD}) describes fully elastic scattering, while the second term is
the recharging term with a pair of neutral pions in the intermediate state converted to $\pi^+\pi^-$ in the final state.
\section{Results}
\begin{figure}[h!]
 \centering
 \includegraphics[scale=0.35,clip]{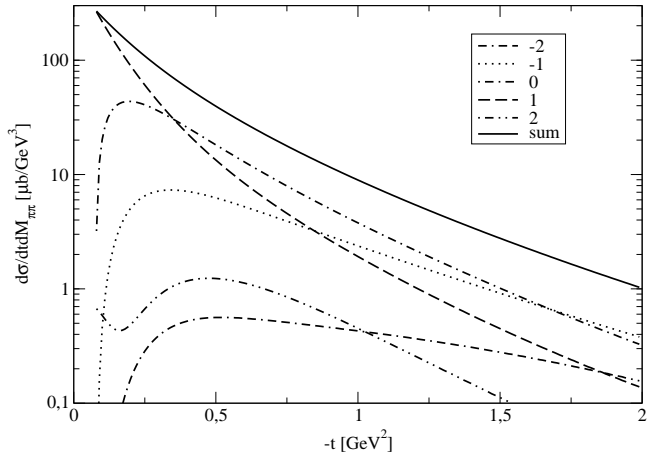}
 \caption{Born cross sections for $\pi^+\pi^-$ photoproduction at $E_\gamma$=3.5 GeV and $M_{\pi\pi}$=1.27 GeV for different 
angular momentum projections (see legend).}
\label{fig:pippim}
\end{figure}

\begin{figure}[h!]
 \centering
 \includegraphics[scale=0.35,clip]{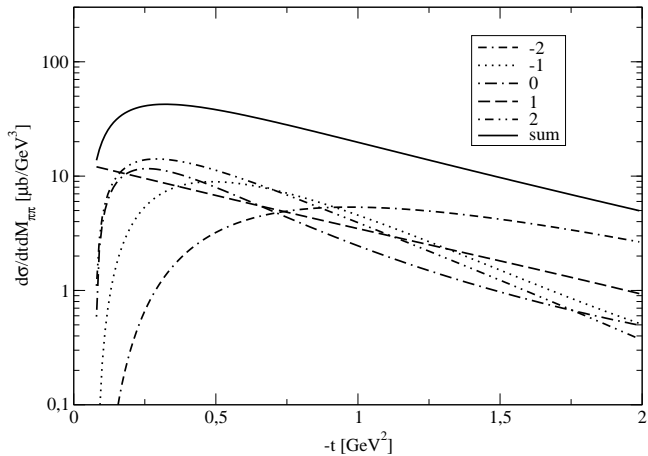}
 \caption{Born cross sections for $\pi^0\pi^0$ photoproduction at $E_\gamma$=3.5 GeV and $M_{\pi\pi}$=1.27 GeV for different 
angular momentum projections (see legend).}
 \label{fig:pi0pi0}
\end{figure}

We have calculated the double differential cross section using the same formula as in \cite{JiKaLe}. Out of 40 spin 
amplitudes describing the $D$-wave $\pi\pi$ photoproduction only 20 are independent due to amplitude invariance under parity 
transformation. So we choose the photon helicity $\lambda_\gamma=+1$ as a reference helicity and refer to amplitudes 
corresponding to various $M$ as no flip, single flip (either up or down), double flip amplitudes and so forth.
From Eq.(\ref{photoamp}) we see that strengths of the photoproduction amplitudes with different $M$ \emph{entirely} depend on 
the Born amplitudes and that final state interactions modulate these amplitudes uniformly. Moreover, the full photoproduction 
amplitude consists of the part proportional to $V_{\pi^+\pi^-}$ and $V_{\pi^0\pi^0}$. So it is interesting to know the
Born cross sections of individual partial waves for both charged and neutral pion pairs. We show these cross sections in Figs. 
\ref{fig:pippim} and \ref{fig:pi0pi0}.
It is worth mentioning that, while $\pi^+\pi^-$ photoproduction is dominated by contributions of $M=$+1,0 and -1 (dashed, 
dot-dashed and dotted curves in Fig.\ref{fig:pippim}), 
$\pi^0\pi^0$ photoproduction has strong contributions of partial waves corresponding to $M=\pm$2 (dot-dot-dashed and 
dash-dash-dotted curves, respectively, in Fig.\ref{fig:pi0pi0}). It can be understood as a 
consequence of double vector meson exchange, as the Born amplitudes for $\pi^0\pi^0$ photoproduction are only type II 
amplitudes. On the other hand, the Born amplitudes for $\pi^+\pi^-$ photoproduction have both type I and type II contributions
with dominating type I contribution.

\begin{figure}[h!]
 \centering
 \includegraphics[scale=.35,clip]{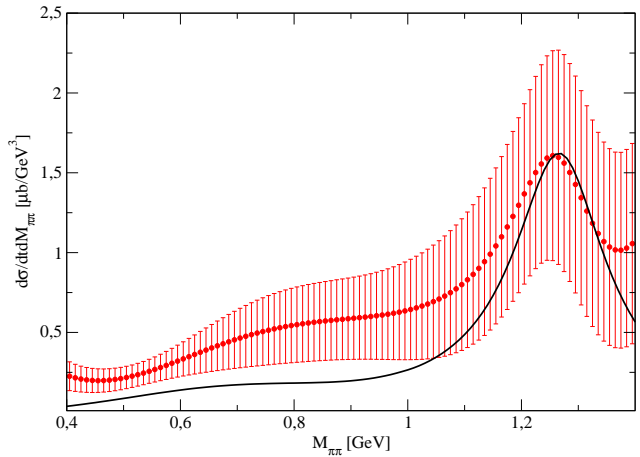}
 \caption{Model prediction for D-wave $\pi^+\pi^-$ mass distribution at $E_\gamma=3.3$ GeV and $-t$=0.55 GeV$^2$ compared to 
CLAS data (color online).}
 \label{figD}
\end{figure}

\begin{figure}[h!]
 \centering
 \includegraphics[scale=.35,clip]{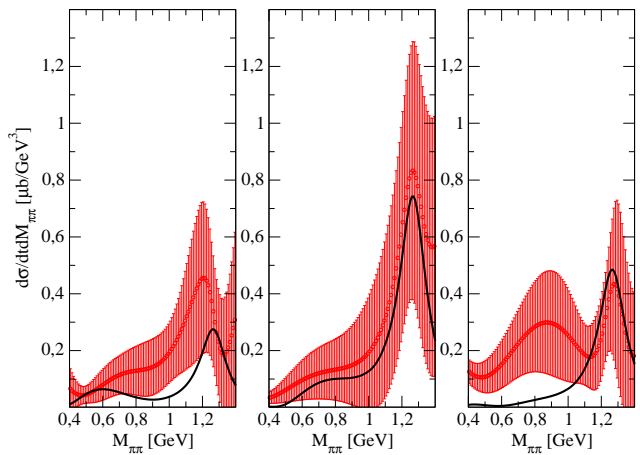}
\caption{Model prediction for $\pi^+\pi^-$ mass distribution for $M$=-1 (left panel), $M$=0 (middle panel) and $M$=+1 (right 
panel) at $E_\gamma=3.3$ GeV and $-t$=0.55 GeV$^2$ compared to CLAS data (color online).}
 \label{figDpm0}
\end{figure}

In Figs. \ref{figD} and \ref{figDpm0}, we show the $D$-wave mass distribution as well as mass distributions for
$M$=-1,0,+1 compared with the corresponding data from CLAS. 
The model quite well reproduces the shape of the resonance. 
The slight asymmetry of the resonance and shift of its maximum towards lower masses observed in the experiment may be attributed to the interference of the resonant $D$-wave 
amplitude with flat contribution of other mechanisms involving pion-nucleon rescattering (Drell mechanism). This feature will 
be accounted for in further studies \cite{BiKa}. Another striking feature of mass distributions corresponding to different values of angular 
momentum projection is that, contrary to vector meson photoproduction where the dominating $M$ coincided with the helicity of 
incident photon (+1 in our convention) in wide range of momentum transfers, $D$-wave photoproduction is dominated by $M=0$ amplitude. 
Our model very well reproduces this feature. Moreover, for $\pi^+\pi^-$ photoproduction it predicts small strengths of the
partial waves corresponding to $M=\pm$2 (they amount to 3.3\% and 1.6\% of the total $D$-wave intensity, respectively). This is in agreement with common practice in experimental analyses, 
where amplitudes with $|M|>$1 are neglected \cite{CLAS,Ballam,Pawlicki}. We stress, however, that this assumption is not true for $\pi^0\pi^0$ 
where contributions of partial waves with $M$=2 are significant.

In actual calculations, we have adopted the definition of \cite{Weinberg} and added the factor $i$ to numerator of the 
propagator (Eq.(\ref{prop})) used in the isoscalar part of the amplitude. 
This reflects the fact that the 
isoscalar amplitude describes the correlated (resonating) pion pair of spin 2. On the other hand, the isotensor amplitude 
describes two uncorrelated pions which essentially propagate independently, thus giving the overall factor of -1. This 
heuristic argument can be substituted in phenomenological applications by introducing an additional correction phase 
between I=0 and I=2 amplitudes and treating it as a model parameter.
\section{Discussion and outlook}
We have presented the theoretical description of the $\pi^+\pi^-$ photoproduction in $D$-wave, treating the resonant behavior
of the amplitude as due to pion-pion final state interactions. In this explanatory study we limited ourselves to the on-shell
part of the amplitude, leaving the analysis of the off-shell effects for further study. $S$-wave analyses suggest 
that off-shell effects can be strong in fact, but proper fixing of the cut-off parameter requires careful fitting to the data and thus 
accounting for other 
mechanisms contributing to the $D$-wave amplitude (like Drell mechanism). This will be the subject of our further study. The model properly 
reproduces relative strengths of different partial waves and, in particular, the fact that the mass distribution in resonance region is 
dominated by the $M$=0 partial wave. Additional check of the model predictions will be the calculation of moments of pion angular distribution
and comparison with moments measured by CLAS experiment.
\begin{acknowledgements}
 This work has been supported by the Polish Ministry of Science and Higher Education (grant No N N202 236940).
\end{acknowledgements}
\appendix*
\section{Matrix elements of the tensor $P_{r,m\overline{m}}^{2M}$ }
In this Appendix, we present the detailed form of the $D$-wave projected elements of the  $P_{r,m\overline{m}}^{2M}$ 
tensor for both type I and type II amplitudes.
The spherical harmonics $\ylm$ used below are understood as $\ylm=Y^l_m(\cos\theta_q,\varphi_q=\pi)$.
This results from the definition of the photon versor by Eq.(\ref{photmom}). In formulas below the off-diagonal, spacial (ie. $i\neq 0$ and $j\neq 0$) tensor elements  are split into nonsymmetric and symmetric parts $P^{2M,ij}_{r}=N^{2M,ij}_{r}+S^{2M,ij}_{r}$ for both 
type I and II amplitudes. For type I amplitudes the tensor components read (we omit the $m\overline{m}$ subscript for brevity):
\begin{widetext}
\begin{equation}
\begin{split}
 P^{2M,x0}_{I}=-\sqrt{2}\sqrt{\frac{4\pi}{3}}\sum_{l=0}^{3}\sum_{m=-l}^{l}\thjr{3}{5}\wthj{1}{0}{l}{0}{2}{0}
\left[\wthj{1}{-1}{l}{m}{2}{M}-\wthj{1}{+1}{l}{m}{2}{M}\right]\ylm Q_l(x),
\end{split}
\end{equation}
\begin{equation}
\begin{split}
 P^{2M,y0}_{I}=-\sqrt{2}i\sqrt{\frac{4\pi}{3}}\sum_{l=0}^{3}\sum_{m=-l}^{l}\thjr{3}{5}\wthj{1}{0}{l}{0}{2}{0}
\left[\wthj{1}{+1}{l}{m}{2}{M}+\wthj{1}{-1}{l}{m}{2}{M}\right]\ylm Q_l(x),
\end{split}
\end{equation}
\begin{equation}
\begin{split}
 P^{2M,z0}_{I}=-2\sqrt{\frac{4\pi}{3}}\sum_{l=0}^{3}\sum_{m=-l}^{l}\thjr{3}{5}\wthj{1}{0}{l}{0}{2}{0}
\wthj{1}{0}{l}{m}{2}{M}\ylm Q_l(x),
\end{split}
\end{equation}
\begin{equation}
 P^{2M,xx}_{I}=P^{2M,xx}_{I;1}+P^{2M,xx}_{I;2},
\end{equation}
where
\begin{equation*}
 \begin{split}
  P^{2M,xx}_{I;1}=-\sqrt{2}\sqrt{\frac{4\pi}{3}}\sum_{l=0}^{3}\sum_{m=-l}^{l}\thjr{3}{5}\wthj{1}{0}{l}{0}{2}{0}
\left[\wthj{1}{-1}{l}{m}{2}{M}-\wthj{1}{+1}{l}{m}{2}{M}\right]\ylm Q_l(x)\hat{q}^x,
 \end{split}
\end{equation*}
\begin{equation*}
 \begin{split}
  P^{2M,xx}_{I;2}&=4\sqrt{\frac{4\pi}{6}}\frac{k}{|\boldsymbol{q}|}\sum_{l=0}^{4}\sum_{m=-l}^{l}
\bigg\{
\sqrt{\frac{2}{3}}
\thjr{1}{5}\wthj{0}{0}{l}{0}{2}{0}\wthj{0}{0}{l}{m}{2}{M}+
\frac{1}{\sqrt{5}}\thjr{5}{5}\wthj{2}{0}{l}{0}{2}{0}\\&
\bigg[
\wthj{2}{-2}{l}{m}{2}{M}+\wthj{2}{+2}{l}{m}{2}{M}-\sqrt{\frac{2}{3}}\wthj{2}{0}{l}{m}{2}{M}\bigg]
\bigg\}\ylm Q_l(x),
 \end{split}
\end{equation*}
\begin{equation}
 \begin{split}
  N^{2M,xy}_{I}=0,
 \end{split}
\end{equation}
\begin{equation}
 \begin{split}
  S^{2M,xy}_{I}=4i\sqrt{\frac{4\pi}{30}}\frac{k}{|\boldsymbol{q}|}
\sum_{l=0}^{4}\sum_{m=-l}^{l}\thjr{5}{5}\wthj{2}{0}{l}{0}{2}{0}\left[\wthj{2}{+2}{l}{m}{2}{M}-\wthj{2}{-2}{l}{m}{2}{M}
\right]\ylm Q_l(x),
 \end{split}
\end{equation}
\begin{equation}
 \begin{split}
  N^{2M,xz}_{I}=-2\sqrt{\frac{4\pi}{3}}\sum_{l=0}^{3}\sum_{m=-l}^{l}\thjr{3}{5}\wthj{1}{0}{l}{0}{2}{0}
\wthj{1}{0}{l}{m}{2}{M}\ylm Q_l(x)\hat{q}^z,
 \end{split}
\end{equation}
\begin{equation}
 \begin{split}
  S^{2M,xz}_{I}=4\sqrt{\frac{4\pi}{30}}\sum_{l=0}^{4}\sum_{m=-l}^{l}\thjr{5}{5}\wthj{2}{0}{l}{0}{2}{0}
\left[
\wthj{2}{-1}{l}{m}{2}{m}-\wthj{2}{+1}{l}{m}{2}{M}
\right]\ylm Q_l(x),
 \end{split}
\end{equation}
\begin{equation}
 \begin{split}
 N^{2M,yx}_{I}=-\sqrt{2}i\sqrt{\frac{4\pi}{3}}\sum_{l=0}^{3}\sum_{m=-l}^{l}\thjr{3}{5}\wthj{1}{0}{l}{0}{2}{0}
\left[\wthj{1}{+1}{l}{m}{2}{M}+\wthj{1}{-1}{l}{m}{2}{M}
\right]\ylm Q_l(x)\hat{q}^x,
 \end{split}
\end{equation}
\begin{equation}
 S^{2M,yx}_{I}=S^{2M,xy}_{I},
\end{equation}
\begin{equation}
 P^{2M,yy}_{I}=P^{2M,yy}_{I;1}+P^{2M,yy}_{I;2},
\end{equation}
where
\begin{equation*}
 P^{2M,yy}_{I;1}=0,
\end{equation*}
\begin{equation*}
 \begin{split}
  P^{2M,yy}_{I;2}&=4\sqrt{\frac{4\pi}{6}}\frac{k}{|\boldsymbol{q}|}\sum_{l=0}^4\sum_{m=-l}^{l}
\bigg\{\sqrt{\frac{2}{3}}\thjr{1}{5}\wthj{0}{0}{l}{0}{2}{0}\wthj{0}{0}{l}{m}{2}{M}
-\frac{1}{\sqrt{5}}\thjr{5}{5}\wthj{2}{0}{l}{0}{2}{0}\\&
\bigg[\wthj{2}{-2}{l}{m}{2}{M}+\wthj{2}{+2}{l}{m}{2}{M} +\sqrt{\frac{2}{3}}\wthj{2}{0}{l}{m}{2}{M}
\bigg]
\bigg\}\ylm Q_l(x),
 \end{split}
\end{equation*}
\begin{equation}
 \begin{split}
  N^{2M,yz}_{I}=-\sqrt{2}i\sqrt{\frac{4\pi}{3}}\sum_{l=0}^3\sum_{m=-l}^l\thjr{3}{5}\wthj{1}{0}{l}{0}{2}{0}& \bigg[
\wthj{1}{+1}{l}{m}{2}{M}+\wthj{1}{-1}{l}{m}{2}{M}\bigg]\ylm Q_l(x)\hat{q}^z,
 \end{split}
\end{equation}
\begin{equation}
 \begin{split}
  S^{2M,yz}_{I}=-4i\sqrt{\frac{4\pi}{30}}\frac{k}{|\boldsymbol{q}|}\sum_{l=0}^4\sum_{m=-l}^l\thjr{5}{5}\wthj{2}{0}{l}{0}{2}{0} \bigg[\wthj{2}{-1}{l}{m}{2}{M}+\wthj{2}{+1}{l}{m}{2}{M},
\bigg]\ylm Q_l(x)
 \end{split}
\end{equation}
\begin{equation}
\begin{split}
 N^{2M,zx}_{I}=-2\sqrt{\frac{4\pi}{3}}\sum_{l=0}^3\sum_{m=-l}^l\thjr{3}{5}\wthj{1}{0}{l}{0}{2}{0}
\wthj{1}{0}{l}{m}{2}{M}\ylm
Q_l(x)\hat{q}^x,
\end{split}
\end{equation}
\begin{equation}
 S^{2M,zx}_{I}=S^{2M,xz}_{I},
\end{equation}
\begin{equation}
 N^{2M,zy}_{I}=0,
\end{equation}
\begin{equation}
 S^{2M,zy}_{I}=S^{2M,yz}_{I},
\end{equation}
\begin{equation}
 P^{2M,zz}_{I}=P^{2M,zz}_{I;1}+P^{2M,zz}_{I;2},
\end{equation}
where
\begin{equation*}
 \begin{split}
  P^{2M,zz}_{I;1}=-2\sqrt{\frac{4\pi}{3}}\sum_{l=0}^{3}\sum_{m=-l}^{l}\thjr{3}{5}\wthj{1}{0}{l}{0}{2}{0}
\wthj{1}{0}{l}{m}{2}{M}\ylm Q_l(x)\hat{q}^z,
 \end{split}
\end{equation*}
\begin{equation*}
 \begin{split}
 P^{2M,zz}_{I;2}=4\frac{\sqrt{4\pi}}{3}\frac{k}{|\boldsymbol{q}|}\sum_{l=0}^4\sum_{m=-l}^{l}
 \bigg\{
 \thjr{1}{5}\wthj{0}{0}{l}{0}{2}{0}\wthj{0}{0}{l}{m}{2}{M}&\\
 +\frac{2}{\sqrt{5}}\thjr{5}{5}\wthj{2}{0}{l}{0}{2}{0}
 \wthj{2}{0}{l}{m}{2}{M}
\bigg\}\ylm Q_l(x),
 \end{split}
\end{equation*}
where $x=\sqrt{1+m_\pi^2/k^2}$ and Legendre functions of the second kind are given by
\begin{equation*}
 Q_0(x)=\frac{1}{2}\ln\frac{x+1}{x-1},
\end{equation*}
\begin{equation*}
 Q_1(x)=\frac{x}{2}\ln\frac{x+1}{x-1}-1,
\end{equation*}
\begin{equation}
 Q_2(x)=\frac{1}{4}(3x^2-1)\ln\frac{x+1}{x-1}-\frac{3}{2}x,
\end{equation}
\begin{equation*}
Q_3(x)=\frac{2}{3}-\frac{5}{2}x^2-\frac{1}{4}x(3-5x^2)\ln\frac{x+1}{x-1},
\end{equation*}
\begin{equation*}
Q_4(x)=\frac{55}{24}x-\frac{35}{8}x^3+\frac{1}{16}(3-30x^2+35x^4)\ln\frac{x+1}{x-1}.
\end{equation*}
For type II amplitudes the corresponding tensor components are
\begin{equation}
 P^{2M,x0}_{II}=P^{2M,x0}_{II;1}+P^{2M,x0}_{II;2},
\end{equation}
where
\begin{equation*}
 \begin{split}
 P^{2M,x0}_{II;1}&=-\sqrt{4\pi}k|\boldsymbol{q}|
 \frac{x}{\sqrt{6}}\sum_{l=0}^3\sum_{m=-l}^l\thjr{3}{5}\wthj{1}{0}{l}{0}{2}{0} \bigg[\wthj{1}{-1}{l}{m}{2}{M}-
 \wthj{1}{+1}{l}{m}{2}{M}\bigg]\ylm Q_l(y),
 \end{split}
\end{equation*}
\begin{equation*}
 \begin{split}
 P^{2M,x0}_{II;2}&=\frac{k|\boldsymbol{q}|}{3}\sqrt{\frac{1}{10}}\sum_{l=0}^4\sum_{m=-l}^l\bigg\{\thjr{5}{5}
 \wthj{2}{0}{l}{0}{2}{0}\bigg[\sqrt{6}\bigg(\rho_{-1}\wthj{2}{-2}{l}{m}{2}{M}-\rho_{+1}\wthj{2}{+2}{l}{m}{2}{M}\bigg)\\&+
 \sqrt{3}\rho_0\bigg(\!\wthj{2}{-1}{l}{m}{2}{M}-\wthj{2}{+1}{l}{m}{2}{M}\!\bigg)+2\rho_{+1}\wthj{2}{0}{l}{m}{2}{M}
\bigg]\\&-2\sqrt{5}\rho_{+1}\thjr{1}{5}\wthj{0}{0}{l}{0}{2}{0}\wthj{0}{0}{l}{m}{2}{M}
 \bigg\}\ylm Q_l(y),
 \end{split}
\end{equation*}
\begin{equation}
P^{2M,y0}_{II}=P^{2M,y0}_{II;1}+P^{2M,y0}_{II;2},
\end{equation}
where
\begin{equation*}
 \begin{split}
 P^{2M,y0}_{II;1}&=-i\sqrt{4\pi}k|\boldsymbol{q}|
 \frac{x}{\sqrt{6}}\sum_{l=0}^3\sum_{m=-l}^l\thjr{3}{5}\wthj{1}{0}{l}{0}{2}{0} \bigg[\wthj{1}{+1}{l}{m}{2}{M}+
 \wthj{1}{-1}{l}{m}{2}{M}\bigg]\ylm Q_l(y),
 \end{split}
\end{equation*}
\begin{equation*}
 \begin{split}
 P^{2M,y0}_{II;2}&=
 \frac{ik|\boldsymbol{q}|}{3}\sqrt{\frac{1}{10}}\sum_{l=0}^4\sum_{m=-l}^l\bigg\{\thjr{5}{5}
 \wthj{2}{0}{l}{0}{2}{0}\\&\big[\sqrt{6}\bigg(\rho_{+1}\wthj{2}{+2}{l}{m}{2}{M}+\rho_{-1}\wthj{2}{-2}{l}{m}{2}{M}\bigg)+
 \sqrt{3}\rho_0\bigg(\wthj{2}{+1}{l}{m}{2}{M}+\wthj{2}{-1}{l}{m}{2}{M}\bigg)
 \bigg\}\ylm Q_l(y),
 \end{split}
\end{equation*}
\begin{equation}
 P^{2M,z0}_{II}=P^{2M,z0}_{II;1}+P^{2M,z0}_{II;2},
\end{equation}
where
\begin{equation*}
 \begin{split}
 P^{2M,z0}_{II;1}&=-\sqrt{4\pi}k|\boldsymbol{q}|
 \frac{x}{\sqrt{3}}\sum_{l=0}^3\sum_{m=-l}^l\thjr{3}{5}\wthj{1}{0}{l}{0}{2}{0} \wthj{1}{0}{l}{m}{2}{M}\ylm Q_l(y),
 \end{split}
\end{equation*}
\begin{equation*}
 \begin{split}
 P^{2M,z0}_{II;2}&=\frac{k|\boldsymbol{q}|}{3}\sqrt{\frac{1}{5}}\sum_{l=0}^4\sum_{m=-l}^l
 \bigg\{
 \thjr{5}{5}\wthj{2}{0}{l}{0}{2}{0}
 \bigg[\sqrt{3}\bigg(\rho_{-1}\wthj{2}{-1}{l}{m}{2}{M}+\rho_{+1}\wthj{2}{+1}{l}{m}{2}{M}\bigg)\\&+2\rho_0
 \wthj{2}{0}{l}{m}{2}{M}\bigg]
 +\thjr{1}{5}\wthj{0}{0}{l}{0}{2}{0}\rho_0\wthj{0}{0}{l}{m}{2}{M}
 \bigg\}\ylm Q_l(y),
 \end{split}
\end{equation*}
\begin{equation}
 \begin{split}
 P^{2M,xx}_{II}=P^{2M,xx}_{II;1}+P^{2M,xx}_{II;2}+P^{2M,xx}_{II;3}+P^{2M,xx}_{II;4}+P^{2M,xx}_{II;5},
 \end{split}
\end{equation}
where
\begin{equation*}
 \begin{split}
 P^{2M,xx}_{II;1}=\sqrt{4\pi}[x(x^2+1)k^2-m_{\pi}^2 x+x^2|\boldsymbol{q}|k]{Y^2_M}^\ast(\hat{q})Q_2(y),
 \end{split}
\end{equation*}
\begin{equation*}
\begin{split}
 P^{2M,xx}_{II;2}&=-\sqrt{\frac{4\pi}{3}}[k^2(x^2+1)+m_{\pi}^2]\sum_{l=0}^3\sum_{m=-l}^l\thjr{3}{5}\wthj{1}{0}{l}{0}{2}{0}\\&
 \bigg[
 \rho_{-1}\wthj{1}{-1}{l}{m}{2}{M}+\rho_0\wthj{1}{0}{l}{m}{2}{M}+\rho_{+1}\wthj{1}{+1}{l}{m}{2}{N}
 \bigg]\ylm Q_l(y),
\end{split}
\end{equation*}
\begin{equation*}
 \begin{split}
 P^{2M,xx}_{II;3}&=-\sqrt{4\pi}\frac{k|\boldsymbol{q}|}{3}
 \Bigg\{
 \sum_{l=0}^4\sum_{m=-l}^l\bigg\{\sqrt{\frac{2}{5}}\thjr{5}{5}\wthj{2}{0}{l}{0}{2}{0}\\&\bigg[
 \sqrt{3}\bigg(\rho_{-1}^2\wthj{2}{-2}{l}{m}{2}{M}+\rho_{+1}^2\wthj{2}{+2}{l}{m}{2}{M}\bigg)
 +\sqrt{6}\rho_0\bigg(\rho_{-1}\wthj{2}{-1}{l}{m}{2}{M}+\rho_{+1}\wthj{2}{+1}{l}{m}{2}{M}\bigg)\\&
 +\sqrt{2}(\rho_0^2+\rho_{-1}\rho_{+1})\wthj{2}{0}{l}{m}{2}{M}
 \bigg]+\thjr{1}{5}\wthj{0}{0}{l}{0}{2}{0}\wthj{0}{0}{l}{m}{2}{M}
 \bigg\}\ylm Q_l(y)
 \Bigg\},
 \end{split}
\end{equation*}
\begin{equation*}
 P^{2M,xx}_{II;4}=P^{2M,xx}_{II;4a}+P^{2M,xx}_{II;4b},
\end{equation*}
 and
\begin{equation*}
 \begin{split}
  P^{2M,xx}_{II;4a}&=k\sqrt{\frac{4\pi}{6}}(2kx^2-|\boldsymbol{q}|x)\sum_{l=0}^3\sum_{m=-l}^l\thjr{3}{5}
 \wthj{1}{0}{l}{0}{2}{0}\bigg[\wthj{1}{-1}{l}{m}{2}{m}-\wthj{1}{+1}{l}{m}{2}{M}\bigg]\ylm Q_l(y)\hat{q}^x,
 \end{split}
\end{equation*}
\begin{equation*}
 \begin{split}
  P^{2M,xx}_{II;4b}&=\sqrt{4\pi}\frac{k|\boldsymbol{q}|}{3\sqrt{2}}
  \Bigg\{\sum_{l=0}^4\sum_{m=-l}^l
  \bigg\{
  \frac{1}{\sqrt{5}}\thjr{5}{5}\wthj{2}{0}{l}{0}{2}{0}\\&
  \bigg[
  \sqrt{6}\bigg(\rho_{-1}\wthj{2}{-2}{l}{m}{2}{M}-\rho_{+1}\wthj{2}{+2}{l}{m}{2}{M}\bigg)+
  \sqrt{3}\rho_0\bigg(\wthj{2}{-1}{l}{m}{2}{M}-\wthj{2}{+1}{l}{m}{2}{M}\bigg)\\&+
  (\rho_{+1}-\rho_{-1})\wthj{2}{0}{l}{m}{2}{M}
  \bigg]
  -(\rho_{+1}-\rho_{-1})\thjr{1}{5}\wthj{0}{0}{l}{m}{2}{0}\wthj{0}{0}{l}{m}{2}{M}
  \bigg\}\ylm Q_l(y)\hat{q}^x
  \Bigg\},
 \end{split}
\end{equation*}
\begin{equation*}
 \begin{split}
 P^{2M,xx}_{II;5}&=-\sqrt{4\pi}xk^2
 \Bigg\{
 \sum_{l=0}^4\sum_{m=-l}^l
 \bigg\{
 \frac{2}{3}\thjr{1}{5}\wthj{0}{0}{l}{0}{2}{0}\wthj{0}{0}{l}{m}{2}{M}\\&
 +\sqrt{\frac{2}{15}}\thjr{5}{5}\wthj{2}{0}{l}{0}{2}{0}
 \bigg[
 \wthj{2}{-2}{l}{m}{2}{M}+\wthj{2}{+2}{l}{m}{2}{M}-\sqrt{\frac{2}{3}}\wthj{2}{0}{l}{m}{2}{M}
 \bigg]
 \bigg\}\ylm Q_l(y)
 \Bigg\},
 \end{split}
\end{equation*}
\begin{equation}
 \begin{split}
 N^{2M,xy}_{II}=0,
 \end{split}
\end{equation}
\begin{equation}
 \begin{split}
 S^{2M,xy}_{II}&=-i\sqrt{4\pi}xk^2\sqrt{\frac{2}{15}}\sum_{l=0}^4\sum_{m=-l}^l\thjr{5}{5}\wthj{2}{0}{l}{0}{2}{0}
 \bigg[\wthj{2}{+2}{l}{m}{2}{M}-\wthj{2}{-2}{l}{m}{2}{M}\bigg]\ylm Q_l(y),
 \end{split}
\end{equation}
\begin{equation}
 N^{2M,xz}_{II}=N^{2M,xz}_{II;1}+N^{2M,xz}_{II;2},
\end{equation}
where
\begin{equation*}
 \begin{split}
 N^{2M,xz}_{II;1}&=\sqrt{\frac{4\pi}{6}}k(2kx^2-|\boldsymbol{q}|x)\lsum{3}\thjr{3}{5}\wthj{1}{0}{l}{0}{2}{0}
 \bigg[
 \wthj{1}{-1}{l}{m}{2}{M}-\wthj{1}{+1}{l}{m}{2}{M}
 \bigg]\ylm Q_l(y)\hat{q}^z,
 \end{split}
\end{equation*}
\begin{equation*}
 \begin{split}
 N^{2M,xz}_{II;2}&=\sqrt{\frac{4\pi}{3}}\frac{k|\boldsymbol{q}|}{\sqrt{6}}
 \Bigg\{
 \lsum{4}
 \bigg\{
 \frac{1}{\sqrt{5}}\thjr{5}{5}\wthj{2}{0}{l}{0}{2}{0}\\&
 \bigg[
 \sqrt{6}\bigg(\rho_{-1}\wthj{2}{-2}{l}{m}{2}{M}-\rho_{+1}\wthj{2}{+2}{l}{m}{2}{M}\bigg)
 +\sqrt{3}\rho_0\bigg(\wthj{2}{-1}{l}{m}{2}{M}-\wthj{2}{+1}{l}{m}{2}{M}\bigg)\\&+(\rho_{+1}-\rho_{-1})\wthj{2}{0}{l}{m}{2}{M}
 \bigg]
 -(\rho_{+1}-\rho_{-1})\thjr{1}{5}\wthj{0}{0}{l}{0}{2}{0}\wthj{0}{0}{l}{m}{2}{M}
 \bigg\}\ylm Q_l(y)
 \Bigg\}\hat{q}^z,
 \end{split}
\end{equation*}
\begin{equation}
 \begin{split}
 S^{2M,xz}_{II}&=-\sqrt{\frac{4\pi}{3}}\sqrt{\frac{2}{5}}xk^2\lsum{4}\thjr{5}{5}\wthj{2}{0}{l}{0}{2}{0}
 \bigg[\wthj{2}{-1}{l}{m}{2}{M}-\wthj{2}{+1}{l}{m}{2}{M}\bigg]\ylm Q_l(y),
 \end{split}
\end{equation}
\begin{equation}
 \begin{split}
 N^{2M,yx}_{II}=N^{2M,yx}_{II;1}+N^{2M,yx}_{II;2},
 \end{split}
\end{equation}
where
\begin{equation*}
 \begin{split}
 N^{2M,yx}_{II;1}&=\sqrt{4\pi}\frac{ik(2kx^2-|\boldsymbol{q}x|)}{\sqrt{6}}\lsum{3}\thjr{3}{5}\wthj{1}{0}{l}{0}{2}{0}
 \bigg[
 \wthj{1}{+1}{l}{m}{2}{M}+\wthj{1}{-1}{l}{m}{2}{M}
 \bigg]\ylm Q_l(y)\hat{q}^x,
 \end{split}
\end{equation*}
\begin{equation*}
 \begin{split}
 N^{2M,yx}_{II;2}&=\sqrt{4\pi}\frac{ik|\boldsymbol{q}|}{\sqrt{30}}\lsum{4}\thjr{5}{5}\wthj{2}{0}{l}{0}{2}{0}\\&
 \bigg[
 \rho_0\bigg(\wthj{2}{+1}{l}{m}{2}{M}+\wthj{2}{-1}{l}{m}{2}{M}\bigg)
 +\sqrt{2}\bigg(\rho_{+1}\wthj{2}{+2}{l}{m}{2}{M}+\rho_{-1}\wthj{2}{-2}{l}{m}{2}{M}\bigg)
 \bigg]\ylm Q_l(y)\hat{q}^x,
 \end{split}
\end{equation*}
\begin{equation}
 \begin{split}
 S^{2M,yx}_{II}=S^{2M,xy}_{II},
 \end{split}
\end{equation}
\begin{equation}
 \begin{split}
 P^{2M,yy}_{II}=P^{2M,yy}_{II;1}+P^{2M,yy}_{II;2}+P^{2M,yy}_{II;3}+P^{2M,yy}_{II;4}+P^{2M,yy}_{II;5},
 \end{split}
\end{equation}
where
\begin{equation*}
 P^{2M,yy}_{II;1}=P^{2M,xx}_{II;1}, \qquad
 P^{2M,yy}_{II;2}=P^{2M,xx}_{II;2}, \qquad
 P^{2M,yy}_{II;3}=P^{2M,xx}_{II;3}, \qquad
\end{equation*}
\begin{equation*}
 P^{2M,yy}_{II;4}=0,
\end{equation*}
\begin{equation*}
 \begin{split}
 P^{2M,yy}_{II;5}&=-\sqrt{4\pi}\sqrt{\frac{2}{3}}xk^2\Bigg\{
 \lsum{4}\bigg\{\sqrt{\frac{2}{3}}\thjr{1}{5}\wthj{0}{0}{l}{0}{2}{0}\wthj{0}{0}{l}{m}{2}{M}\\&
 -\frac{1}{\sqrt{5}}\thjr{5}{5}\wthj{2}{0}{l}{0}{2}{0}
 \bigg[
 \wthj{2}{-2}{l}{m}{2}{M}+\wthj{2}{+2}{l}{m}{2}{M}+\sqrt{\frac{2}{3}}\wthj{2}{0}{l}{m}{2}{M}
 \bigg]
 \bigg\}\ylm Q_l(y)
 \Bigg\},
 \end{split}
\end{equation*}
\begin{equation}
 N^{2M,yz}_{II}=N^{2M,yz}_{II;1}+N^{2M,yz}_{II;2},
\end{equation}
where
\begin{equation*}
 \begin{split}
 N^{2M,yz}_{II;1}&=\sqrt{4\pi}\frac{ik(2kx^2-|\boldsymbol{q}x|)}{\sqrt{6}}\lsum{3}\thjr{3}{5}\wthj{1}{0}{l}{0}{2}{0}
 \bigg[
 \wthj{1}{+1}{l}{m}{2}{M}+\wthj{1}{-1}{l}{m}{2}{M}
 \bigg]\ylm Q_l(y)\hat{q}^z,
 \end{split}
\end{equation*}
\begin{equation*}
 \begin{split}
 N^{2M,yz}_{II;2}&=\sqrt{4\pi}\frac{ik|\boldsymbol{q}|}{\sqrt{30}}\lsum{4}\thjr{5}{5}\wthj{2}{0}{l}{0}{2}{0}\\&
 \bigg[
 \rho_0\bigg(\wthj{2}{+1}{l}{m}{2}{M}+\wthj{2}{-1}{l}{m}{2}{M}\bigg)
 +\sqrt{2}\bigg(\rho_{+1}\wthj{2}{+2}{l}{m}{2}{M}+\rho_{-1}\wthj{2}{-2}{l}{m}{2}{M}\bigg)
 \bigg]\ylm Q_l(y)\hat{q}^z,
 \end{split}
\end{equation*}
\begin{equation}
 \begin{split}
 S^{2M,yz}_{II}&=\sqrt{\frac{4\pi}{3}}\sqrt{\frac{2}{5}}ik^2x\lsum{4}\thjr{5}{5}\wthj{2}{0}{l}{0}{2}{0}
 \bigg[
 \wthj{2}{-1}{l}{m}{2}{M}+\wthj{2}{+1}{l}{m}{2}{M}
 \bigg]\ylm Q_l(y),
 \end{split}
\end{equation}
\begin{equation}
 N^{2M,zx}_{II}=N^{2M,zx}_{II;1}+N^{2M,zx}_{II;2},
\end{equation}
where
\begin{equation*}
 \begin{split}
 N^{2M,zx}_{II;1}&=\sqrt{\frac{4\pi}{3}}k(2kx^2-|\boldsymbol{q}|x)\lsum{3}\thjr{3}{5}\wthj{1}{0}{l}{0}{2}{0}
 \wthj{1}{0}{l}{m}{2}{M}\ylm Q_l(y)\hat{q}^x,
 \end{split}
\end{equation*}
\begin{equation*}
 \begin{split}
 N^{2M,zx}_{II;2}&=\frac{\sqrt{4\pi}}{3}k|\boldsymbol{q}|
 \Bigg\{
 \lsum{4}\bigg\{
 \frac{1}{\sqrt{5}}\thjr{5}{5}\wthj{2}{0}{l}{0}{2}{0}\\&
 \bigg[
 \sqrt{3}\bigg(\rho_{-1}\wthj{2}{-1}{l}{m}{2}{M}+\rho_{+1}\wthj{2}{+1}{l}{m}{2}{M}\bigg)
 +2\rho_0\wthj{2}{0}{l}{m}{2}{M}
 \bigg]\\&
 +\rho_0\thjr{1}{5}\wthj{0}{0}{l}{0}{2}{0}
 \bigg\}\ylm Q_l(y)
 \Bigg\}\hat{q}^x,
 \end{split}
\end{equation*}
\begin{equation}
 \begin{split}
 S^{2M,zx}_{II}=S^{2M,xz}_{II},
 \end{split}
\end{equation}
\begin{equation}
 \begin{split}
 N^{2M,zy}_{II}=0,
 \end{split}
\end{equation}
\begin{equation}
 \begin{split}
 S^{2M,zy}_{II}=S^{2M,yz}_{II},
 \end{split}
\end{equation}
\begin{equation}
 P^{2M,zz}_{II}=P^{2M,zz}_{II;1}+P^{2M,zz}_{II;2}+P^{2M,zz}_{II;3}+P^{2M,zz}_{II;4}+P^{2M,zz}_{II;5},
\end{equation}
where
\begin{equation*}
 P^{2M,zz}_{II;1}=P^{2M,xx}_{II;1}, \qquad
 P^{2M,zz}_{II;2}=P^{2M,xx}_{II;2}, \qquad
 P^{2M,zz}_{II;3}=P^{2M,xx}_{II;3},
\end{equation*}
\begin{equation*}
 P^{2M,zz}_{II;4}=P^{2M,zz}_{II;4a}+P^{2M,zz}_{II;4b},
\end{equation*}
\begin{equation*}
 \begin{split}
  P^{2M,zz}_{II;4a}&=\sqrt{\frac{4\pi}{3}}k(2kx^2-|\boldsymbol{q}|x)\lsum{3}\thjr{3}{5}\wthj{1}{0}{l}{0}{2}{0}
  \wthj{1}{0}{l}{m}{2}{M}\ylm Q_l(y)\hat{q}^z,
 \end{split}
\end{equation*}
\begin{equation*}
 \begin{split}
 P^{2M,zz}_{II;4b}&=\frac{\sqrt{4\pi}}{3}k|\boldsymbol{q}|
 \Bigg\{
 \lsum{4}
 \bigg\{
 \frac{1}{\sqrt{5}}\thjr{5}{5}\wthj{2}{0}{l}{0}{2}{0}\\&
 \bigg[
 \sqrt{3}\bigg(\rho_{-1}\wthj{2}{-1}{l}{m}{2}{M}+\rho_{+1}\wthj{2}{+1}{l}{m}{2}{M}\bigg)+2\rho_0\wthj{2}{0}{l}{m}{2}{M}
 \bigg]\\&
 +\rho_0\thjr{1}{5}\wthj{0}{0}{l}{0}{2}{0}\wthj{0}{0}{l}{m}{2}{M}
 \bigg\}\ylm Q_l(y)
 \Bigg\}\hat{q}^z,
 \end{split}
\end{equation*}
\begin{equation*}
 \begin{split}
  P^{2M,zz}_{II;5}&=-\sqrt{4\pi}\frac{2}{3}xk^2\Bigg\{\lsum{4}\bigg\{
  \thjr{1}{5}\wthj{0}{0}{l}{0}{2}{0}\wthj{0}{0}{l}{m}{2}{M}\\&
  +\frac{2}{\sqrt{5}}\thjr{5}{5}\wthj{2}{0}{l}{0}{2}{0}\wthj{2}{0}{l}{m}{2}{M}
  \bigg\}\ylm Q_l(y)
  \Bigg\}.
 \end{split}
\end{equation*}
\end{widetext}
where for particle $a$ exchanged in the upper part of diagrams shown in Fig. \ref{fig:born} 
\begin{equation}
  y=x+\frac{m_a^2-m_\pi^2}{2|\boldsymbol{q}|k}.
\end{equation}
The coefficients $\rho_{-1}$, $\rho_0$ and $\rho_{-1}$ are used in the expansion of the product $\hat{q}\cdot\hat{\kappa}$
in terms of spherical harmonics:
\begin{equation}
 \hat{q}\cdot\hat{\kappa}=\sqrt{\frac{4\pi}{3}}\sum_{m=-1}^{+1}\rho_m Y^1_m(\Omega)
\end{equation}
and can be expressed by the angle between photon momentum and spin quantisation axis as $\rho_0=\cos\theta_q$, 
$\rho_{\pm 1}=\pm \sin\theta_q/\sqrt{2}$.

\end{document}